\documentclass[aps,nofootinbib,twocolumn,superscriptaddress]{revtex4-1}
\pdfoutput=1
\usepackage{float,graphicx,multirow,bm,color,amsmath,amssymb,hyperref}
\hypersetup{
    colorlinks=true,
    linkcolor=black,
    citecolor=black,
}

\usepackage{color}
\usepackage{comment}
\usepackage{graphicx}
\usepackage{amsmath}
\usepackage{hyperref}

\begin{document}
\title{Impact of theoretical priors in cosmological analyses:\\ the case of single field quintessence}

\author{Simone Peirone}\affiliation{Institute Lorentz, Leiden University, PO Box 9506, Leiden 2300 RA, The Netherlands}
\author{Matteo Martinelli}\affiliation{Institute Lorentz, Leiden University, PO Box 9506, Leiden 2300 RA, The Netherlands}
\author{Marco Raveri}\affiliation{Kavli Institute for Cosmological Physics, Enrico Fermi Institute, The University of Chicago, Chicago, Illinois 60637, USA}
\affiliation{Institute Lorentz, Leiden University, PO Box 9506, Leiden 2300 RA, The Netherlands}
\author{Alessandra Silvestri}\affiliation{Institute Lorentz, Leiden University, PO Box 9506, Leiden 2300 RA, The Netherlands}

\begin{abstract}
We investigate the impact of general conditions of theoretical stability and cosmological viability on dynamical dark energy models. As a powerful example, we study whether minimally coupled, single field Quintessence models that are safe from ghost instabilities, can source the CPL expansion history  recently shown to be mildly favored by a combination of CMB (Planck) and Weak Lensing (KiDS) data.  We find that in their most conservative form, the theoretical conditions impact the analysis in such a way that smooth single field Quintessence becomes significantly disfavored with respect to the standard $\Lambda$CDM cosmological model. This is due to the fact that these conditions cut a significant portion of the $(w_0,w_a)$ parameter space for CPL, in particular eliminating the region that would be favored by weak lensing data. Within the scenario of a smooth dynamical dark energy parametrized with CPL, weak lensing data favors a region that would require multiple fields to ensure gravitational stability.
\end{abstract}

\maketitle

\section{Introduction}\label{sec:intro}
Recent observational results from weak lensing experiments~\citep{Kitching:2014dtq,Hildebrandt:2016iqg}
have been found to be in discordance with cosmic microwave background
measurements from Planck~\citep{Adam:2015rua}. This discordance is usually quantified by means of consistency tests~\cite{Raveri:2015maa,Seehars:2015qza,Grandis:2015qaa,Joudaki:2016mvz} or in terms of the  $S_8 = \sigma_8 \sqrt{\Omega_m /0.3}$ derived parameter, with $\Omega_m$ quantifying the matter density and $\sigma_8$ the amplitude of the linear matter power spectrum on $8\ h^{-1}$ Mpc scales, to which weak lensing surveys are expected to be particularly sensitive~\citep{Joudaki:2016kym}. In the case of the weak lensing Kilo Degree Survey (KiDS), this discordance 
is measured to be at the level of $2.3 \sigma$~\citep{Hildebrandt:2016iqg}.

Despite significant effort in investigating the effects 
of systematics on the measurements of both weak lensing~\citep{Hildebrandt:2016iqg, 
Joudaki:2016kym,Joudaki:2016mvz} and Planck~\citep{Aghanim:2015xee}, no  relevant reduction of the tension has been found so far\footnote{We point out that other recent analysis of the KiDS data~ \citep{Kohlinger:2017sxk,vanUitert:2017ieu} yield different degrees of tension with Planck. Our results, however, would not be qualitatively affected by these.}, prompting some initial investigations on whether this discordance
could be due to the assumption of the standard cosmological model, $\Lambda$CDM~\citep{Hamann:2013iba,DiValentino:2015ola}. 
In~\citep{Joudaki:2016kym}, the KiDS collaboration revisited the tensions between the data sets within some  simple extensions of the $\Lambda$CDM model. They found a dynamical dark energy (DE) with a time-varying equation of state to alleviate the tension, reducing it to $S_8$ to $0.91 \sigma$ on $S_8$, \emph{and} to be favored by the combined KiDS and Planck datasets, from a 
bayesian model selection point of view~\citep{Joudaki:2016kym}. 

Given this interesting result, in this paper we investigate how connecting \emph{theoretically viable} DE model with the phenomenological 
parametrization used to perform this analysis will affect the results.  In practice, we ask ourselves which \emph{theoretically viable} DE model could correspond to the expansion history preferred by the combination of KiDS and Planck data. We shall show that the implementation of conditions of theoretical consistency inside the analysis pipeline, in the form of viability priors, allows to answer such questions in a straighforward way; more generally, it allows to perform theoretically informed data analysis in a very efficient way. For a previous investigation of the impact of prior information in quintessence studies see~\cite{Marsh:2014xoa}.

The paper is organized as follows. In Section~\ref{sec:stab} we revisit general conditions of stability for DE models, focusing on minimally coupled quintessence. We then describe how we implement these conditions in our exploration of CPL models. In Section~\ref{sec:data} we describe the data sets that we employ and, finally, we show the results of our analysis in Section~\ref{sec:res}. We draw our Conclusions
in Section~\ref{sec:conc}.

\section{Dynamical dark energy: stability and viability conditions}\label{sec:stab}
\begin{figure*}
 \begin{center}
   \includegraphics[width=0.8\textwidth]{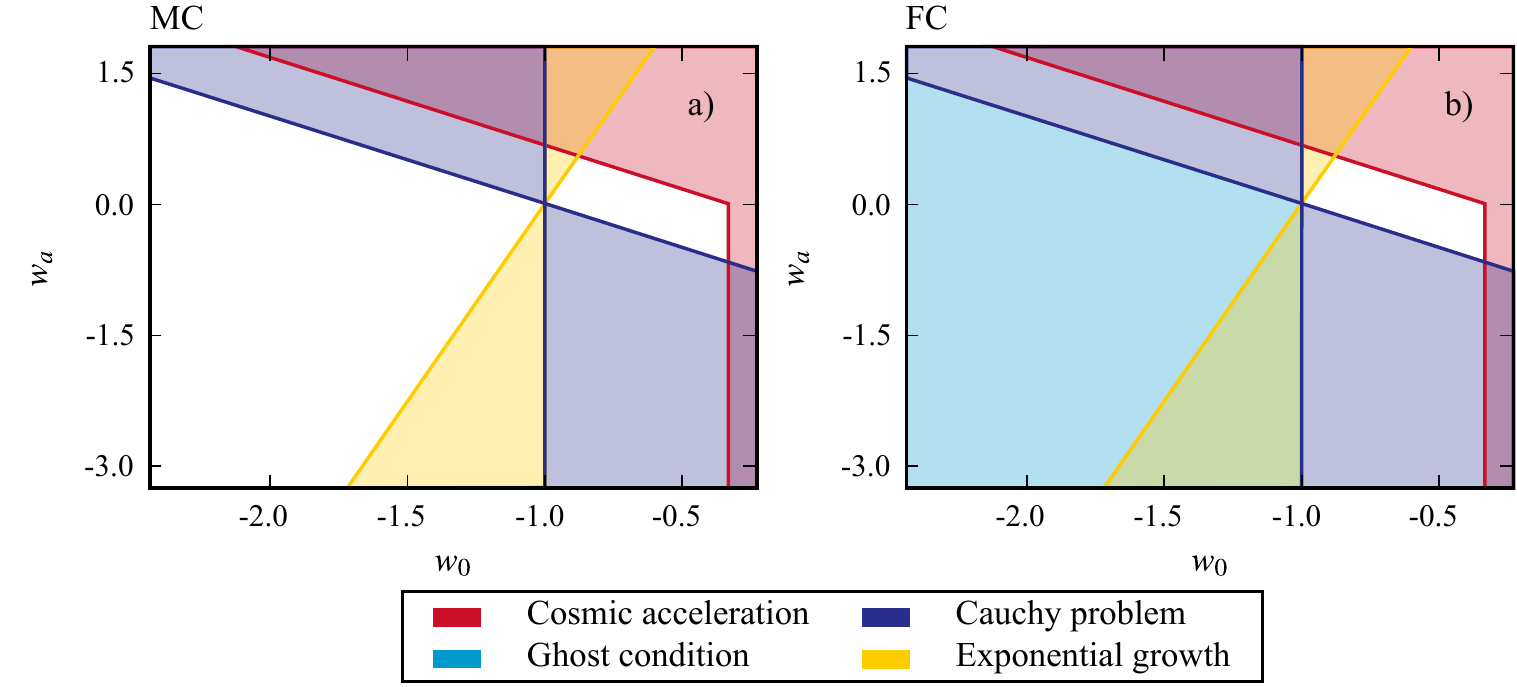}
  \caption{Viable and non-accessible regions of the CPL parameter space within the realm of \emph{standard quintessence}. Different colours correspond to different viability and stability conditions, as shown in legend. 
  In the \emph{left} panel we show the effects of Mathematical Conditions (MC)
  and in the \emph{right} one, we show the full combination (FC) of mathematical and physical conditions.   }
  \label{fig:stability}
 \end{center}
\end{figure*}
In the analysis of observational data it is often useful to explore broad classes of models through a framework that parametrizes the relevant deviations from the standard $\Lambda$CDM cosmological model.
This approach is commonly applied to the investigation of dynamical DE models, where the equation of state  is  often described by the  Chevallier-Polarski-Linder parametrization~\cite{Chevallier:2000qy,Linder:2002et}:
\begin{equation} \label{CPL}
w_{\rm DE}(a) = w_0 + w_a (1-a).
\end{equation}
When going beyond background probes, the latter is commonly combined with  the Parametrized Post-Friedmann (PPF) framework for DE perturbations, introduced in~\cite{Hu:2007pj,Hu:2008zd,Fang:2008sn} and implemented in the Einstein-Boltzmann solver \texttt{CAMB} ~\cite{Fang:2008sn} .
This offers a stable and accurate description of DE perturbations over the entire parameter space of CPL, provided that the DE field remains smooth with respect to matter on the scales of interest. This is the approach used also in~\citep{Joudaki:2016kym}, where they found a best fit point  in the 3rd quadrant of the $(w_0,w_a)$ plane, corresponding to a $w$ that was below $-1$ in the recent past. 

While adopting a parametrized framework, it is informative to make contact with known theories and associate different regions of the parameter space to different viable classes of theories. 
It is known that a single field, minimally coupled quintessence has to have $w > -1$ in order to be stable. When adopting the CPL parameterization, its (w0, wa) are therefore restricted to regions corresponding to $w > -1$.
Outside of that range, results need to be interpreted within the realm of multifield DE. In the following, we will include these theoretical priors in the statistical analysis pipeline; this will allow us to perform parameter estimation and model selection authomatically accounting for the restrictions that theoretical considerations apply on the parameter space volume.

Minimally coupled, single field models with a speed of sound equal to unity, are arguably the simplest models of dynamical DE. We will refer to them as \emph{standard quintessence}. For these models, the DE field remains smooth on sub-horizon scales, and PPF offers an accurate prescription. However, in this case one should be careful with the allowed regions of the parameter space $(w_0,w_a)$. Indeed, these models would generally suffer from ghost instabilities for   $w<-1$~\cite{Creminelli:2009JCAP,2005PhRvD..71b3515V,2005PhRvD..71d7301H,2005PhRvD..72d3527C}. These instabilities arise from the wrong sign of the kinetic term, which translates into an Hamiltonian unbounded from below and, thus, into an unstable quantum vacuum. Correspondingly, $w=-1$ is referred to as the \emph{phantom divide} and single field models crossing through it are gravitationally unstable.

As explored in~\citep{Carroll:2003PhRvD}, while it is difficult,  it is not impossible to have a quintessence model with $w<-1$ which is \emph{effectively} stable, i.e. the rate of the instabilities is longer than the time scale of interest for the analysis. We keep this option open in our analysis, however we will see that requiring that no instabilities develop over the relevant time interval will still cut a significant portion of the $w<-1$ region.

Alternatively, single field DE models could safely cross the phantom divide if the DE field is non minimally coupled to gravity~\citep{2005PhRvD..71b3525C}, there is kinetically braiding, i.e. mixing of the kinetic terms of the metric and the scalar~\cite{Easson:2016klq,Deffayet:2010qz}, or the model includes higher order derivative operators, as discussed in~\citep{Creminelli:2009JCAP}. 
However, in the former two cases,  the DE component would be clustering and in the latter case, as shown in~\citep{Creminelli:2009JCAP},  stability requires that in the region $w<-1$ the DE field behaves like a k-essence fluid with an approximately zero speed of sound.  
 As such, in all these cases, the DE component cannot be considered smooth on the scales of relevance for large scale structure surveys, and hence the PPF framework does not apply.
Which models could then correspond to a DE which gives a CPL history which crosses the phantom divide, while remaining relatively smooth with respect to matter, (so that PPF is an accurate prescription)? As discussed in~\citep{Hu:2004kh,Guo:2004fq}, one necessarily needs to dwell into the multifield scenarios, with  additional degrees of freedom ensuring gravitational stability. 
This  is the assumption at the heart of the PPF approach.  

In this paper we shall asses the importance of  stability requirements and their impact on the results obtained with the CPL parametrization, by  performing the same  analysis of~\citep{Joudaki:2016kym} within the realm of standard (single field) quintessence. This will allow us to show that, as expected, the model favored by data in~\citep{Joudaki:2016kym} cannot correspond to such a scenario, and more importantly, if one restricts the CPL parameter space to this class of models, then $\Lambda$CDM remains a better fit to the combined data. Thus the result of~\citep{Joudaki:2016kym} should be interpreted in terms of multiple fields DE, highlighting the relevance of sound theoretical conditions when connecting viable models to purely phenomenological approaches.
 
 To this extent,  we do not employ the  PPF approach of CAMB, but rather use the minimal  implementation of CPL in \texttt{EFTCAMB}~\cite{Hu:2013twa,Raveri:2014cka}, which corresponds to a designer standard quintessence with CPL expansion history. 
\texttt{EFTCAMB} has a stability module which imposes two sets of theoretically motivated conditions conditions: Mathematical stability 
Conditions (MC)  and Physical stability Conditions (PC). 
The former are general conditions which do not rely on any assumption connected to a specific theoretical model, but rather ensure mathematical 
consistency and numerical stability of perturbations in the DE sector. 
They include conditions for: the well-posedness of the scalar field perturbations initial value problem;  the avoidance of exponential growth of the perturbations; the existence of a viable cosmological background with matter and radiation eras  and an accelerated phase ($w_{\rm DE}< -1/3$).

On the other hand, PC enforce the absence of ghost and gradient instabilities. Since we put ourselves in the case of a CPL standard quintessence, the stability check will automatically ensure that we do not explore regions of $(w_0,w_a)$ that would correspond to an unstable model. 

While MC and PC can be turned on and off independently, in our analysis we use either MC or the full combination of the two (FC), in order to provide a full protection against instabilities. We stress that a more complete set of PC, e.g. including no-tachyon conditions~\citep{Frusciante2016JCAP}, is being worked out.

In Fig.~\ref{fig:stability} we show the effect of MC and FC on the CPL parameter space, $(w_0,w_a)$.
As we can see in Panel a) MC prohibit the crossing of $w=-1$, cutting a relevant portion of parameter space.
On the other hand, as shown in Panel b), the only cut introduced by PC is the one coming from the no-ghost condition, which cuts the portion of the parameter space corresponding to $w<-1$ at all times. 
Comparing the two panels, we can see that MC allow some parts of the $w<-1$ region, namely those for which the instabilities are still there (from the theoretical point of view), but do not affect the observables, i.e. they do not develop over the relevant time range. 

In this paper we compare our results for CPL quintessence under MC and FC, with those of~\citep{Joudaki:2016kym}, where the PPF module was used.  The results of our analysis
will show how much impact MC and FC can have on the final results and will serve us as an example to stress the power of theoretical stability and viability  conditions in the analysis of cosmological data.

\section{Data analysis}\label{sec:data}
Following~\cite{Joudaki:2016kym}, we consider the full set of data from the tomographic weak gravitational lensing analysis of $\sim450$deg$^2$ by the four-band imaging Kilo Degree Survey (KiDS)~\cite{2013ExA....35...25D,Kuijken:2015vca,Hildebrandt:2016iqg}, , including baryonic effects in the nonlinear matter power spectrum with HMCODE~\cite{2015MNRAS.454.1958M}.
In addition, we also consider the Planck measurements~\cite{Ade:2015xua,Aghanim:2015xee} of CMB temperature and polarization on large angular scales, limited to multipoles $\ell\le 29$ (low-$\ell$ TEB likelihood) and the CMB temperature on smaller angular scales (PLIK TT likelihood).\footnote{Notice that the results would change slightly if the re-analysis of Planck data performed in \citep{Aghanim:2016yuo} was used instead of the 2015 release. This dataset would lower the significance of the discordance between KiDS and Planck results, although not affecting the considerations made in this paper.}

We use \texttt{EFTCAMB} and \texttt{EFTCosmoMC}~\cite{Hu:2013twa,Raveri:2014cka} patches of \texttt{CAMB}/\texttt{CosmoMC} codes~\cite{Lewis:1999bs,Lewis:2002ah}, and we implement these in the version of \texttt{CosmoMC} made publicly available by the KiDS collaboration~\cite{Joudaki:2016kym}.
We analyze KiDS and Planck both separately and combining them, sampling the standard cosmological parameters, i.e. the baryon and cold dark matter energy densities $\Omega_b h^2$ and $\Omega_c h^2$, the optical depth at reionization $\tau$, the amplitude and tilt of primordial power spectrum $\ln{10^{10}A_s}$ and $n_s$ and the  the angular size of the sound horizon at last scattering surface $\theta$. 
To these we add the CPL parameters $w_0$ and $w_a$ when we analyze the extension to the $\Lambda$CDM model, and we perform this analysis both in the MC and FC cases.

We adopt flat priors on the sampled parameters, using the same prior ranges defined in Table 1 of~\cite{Joudaki:2016kym}.

In order to assess the effect of the conditions on the possibility of reducing the low-high redshift tension with a dynamical Dark Energy, we exploit the same statistical tools used in \cite{Joudaki:2016kym}; this will allow us to compare our results to the standard PPF treatment of a dynamical Dark Energy.
We therefore consider the tension between Planck (P) and KiDS (K) datasets on the $S_8=\sigma_8\sqrt{\Omega_m/0.3}$ parameter, defining it as
\begin{equation}
 T(S_8)=\frac{\left|S_8^{P}-S_8^{K}\right|}{\sqrt{\sigma^2(S_8^{P})+\sigma^2(S_8^{K})}} \,.
\end{equation}
We also assess the preference of the data for any of the considered models over $\Lambda$CDM computing the {\it Deviance Information Criterion} (DIC) \cite{RSSB:RSSB12062}:
\begin{equation}
\text{DIC}\equiv\chi^2_{\rm eff}(\hat{\theta})+2p_D \,,
\end{equation}
with $\chi^2_{\rm eff}(\hat{\theta})= -2\ln{\mathcal{L}(\hat{\theta})}$, $\hat{\theta}$ the parameters vector at the maximum likelihood and $p_D=\overline{\chi^2_{\rm eff}(\theta)}-\chi^2_{\rm eff}(\hat{\theta})$, where the bar denotes the average taken over the posterior distribution.
The DIC accounts both for the goodness of fit through $\chi^2_{\rm eff}(\hat{\theta})$ and for the bayesian complexity of the model, $p_D$, which disfavours more complex models.
When comparing $\Lambda$CDM with its extension (ext), we compute:
\begin{equation}
\Delta\text{DIC} = \text{DIC}_{\text{ext}} - \text{DIC}_{\Lambda \text{CDM}}.
\end{equation} 
From this definition it follows that a negative $\Delta\text{DIC}$ would support the extended model, while a positive one would support $\Lambda$CDM.
Finally, we exploit the DIC to assess the concordance of the two different datasets as~\cite{Joudaki:2016mvz}
\begin{equation}
 \log{\mathcal{I}}=-\frac{\mathcal{G}(P,K)}{2},
\end{equation}
with $\mathcal{G}$ defined as
\begin{equation}
 \mathcal{G}(P,K)\equiv \text{DIC}(P \cup K) -\text{DIC}(P)-\text{DIC}(K),
\end{equation}
where $P \cup K$ denotes the combination of Planck and KiDS datasets.
\begin{figure*}
 \begin{center}
   \includegraphics[width=0.8\textwidth]{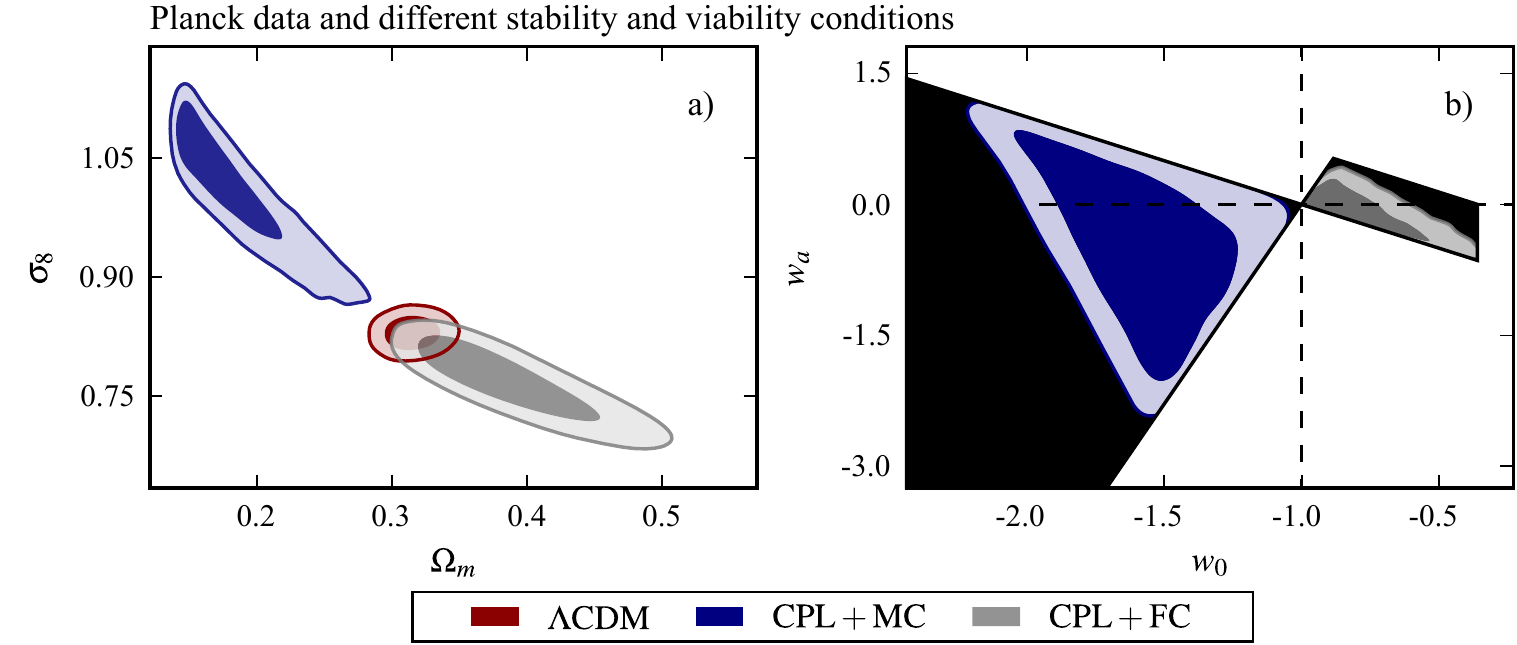}
     \caption{ 
     The joint marginalized posterior of $\Omega_m$ and $\sigma_8$ (panel a) and $w_0$ and $w_a$ (panel b) as obtained with the Planck data in $\Lambda$CDM (red contours), CPL with MC (blue contours) and with FC (gray contours).
  The darker and lighter shades correspond respectively to the $68\%$ C.L. and the $95\%$ C.L. regions. The dashed line indicates the point corresponding to the $\Lambda$CDM model in the $w_0-w_a$ plane.
  }
  \label{fig:allstabres}
 \end{center}
\end{figure*}
\section{Results}\label{sec:res}

 In this section we report the results obtained with the analysis described in section~\ref{sec:data}. We do not 
report the results on all the cosmological parameters, but rather focus on the quantities introduced above
to quantify the tension between the datasets and to assess the effects of the theoretical conditions on it.

The left panel of Fig.~\ref{fig:allstabres} shows the constraints obtained in the $\sigma_8-\Omega_m$ plane analyzing Planck data assuming $\Lambda$CDM and CPL quintessence with the two possible conditions applied to the model. 

We notice that applying MC qualitatively preserves the behavior of the CPL analysis with the PPF approach \cite{Joudaki:2016kym}, allowing for smaller values of $\Omega_m$.
Adopting the FC instead, moves the contours toward higher $\Omega_m$ values.
This difference is due to the change in the allowed $w_{\rm DE}$ values, as can be seen in the right panel of Figure~\ref{fig:allstabres}.
The analysis with MC disfavors the $w_{\rm DE}>-1.0$ region, while the analysis with FC is constrained to be in the $w_{\rm DE}>-1.0$ region.
Secondly, we see that the FC are much more effective in constraining the $w_a$ parameter as the geometric degeneracy between $w_0$ and $w_a$ is broken by viability conditions.
This leads to tighter constraints also in the $\sigma_8$-$\Omega_m$ plane.

\begin{figure*}
 \begin{center}
   \includegraphics[width=0.8\textwidth]{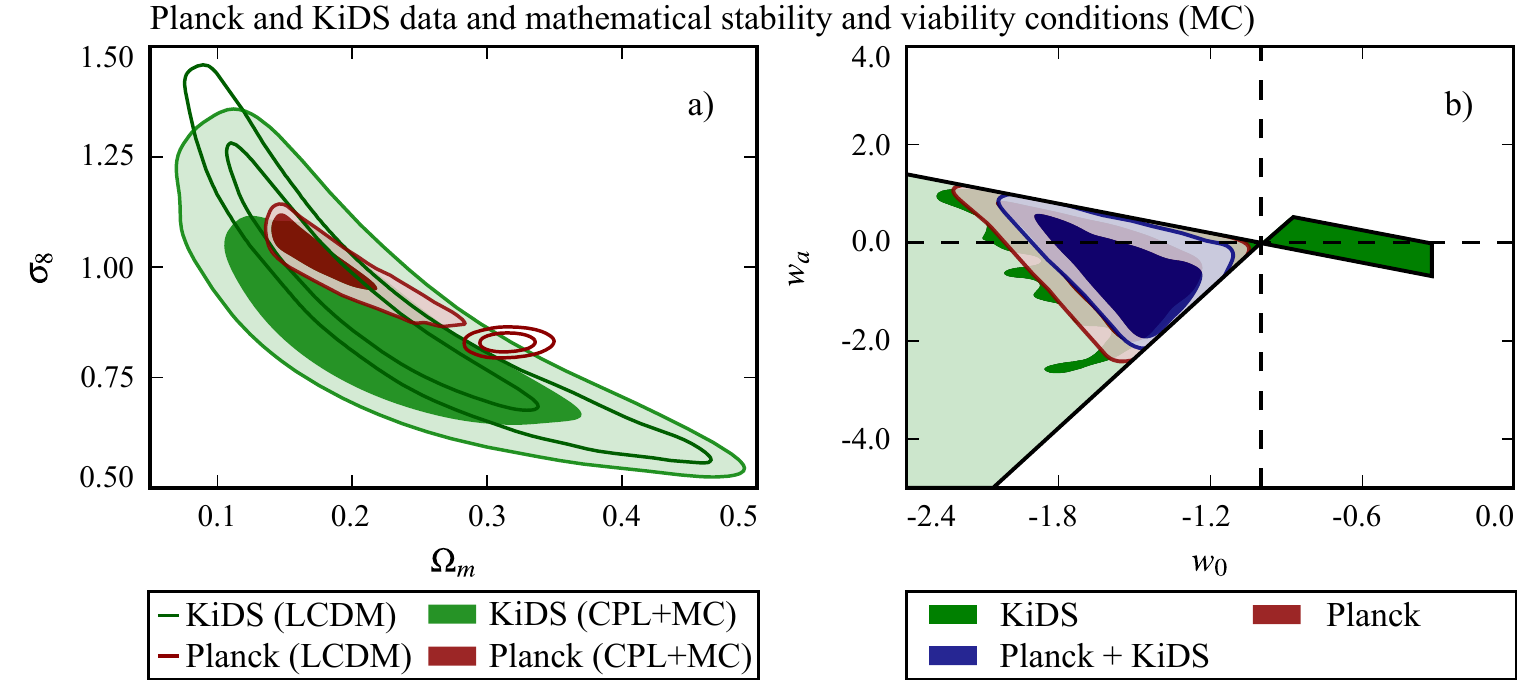}
  \caption{
  The joint marginalized posterior of $\Omega_m$ and $\sigma_8$ (panel a) and $w_0$ and $w_a$ (panel b) as obtained analyzing KiDS (green contours) and Planck (red contours) data. Filled contours refer to constraints obtained in CPL with MC, while empty contours (left panel only) refer to $\Lambda$CDM constraints.
  The darker and lighter shades correspond respectively to the $68\%$ C.L. and the $95\%$ C.L. regions. The dashed line indicates the point corresponding to the $\Lambda$CDM model in the $w_0-w_a$ plane.}

  \label{fig:mathres}
 \end{center}
\end{figure*}

Given this behavior, we expect the CPL quintessence, with MC, to preserve the ability to ease the tension between Planck and KiDS data. 
This is indeed the case, as can be seen in the left panel of Figure~\ref{fig:mathres} and from Table~\ref{tab:tension}.
Following the hierarchy discussed in~\cite{Joudaki:2016kym}, the values $T(S_8)=1$ and $\log{\mathcal{I}}=0.97$ highlight how the tension is removed and the two datasets are now in substantial concordance. 
The right panel of Figure~\ref{fig:mathres} shows the constraints on $w_0$ and $w_a$. Notice that when the two datasets are combined a deviation of more than $2\sigma$ from the $\Lambda$CDM model is found.
Computing then $\Delta$DIC results in a moderate preference of the data for the CPL model when combining Planck and KiDS (see Table~\ref{tab:modsel}).
As expected these results are in agreement with those obtained using a PPF approach~\cite{Joudaki:2016kym}.

\begin{figure*}
 \begin{center}
	\includegraphics[width=0.8\textwidth]{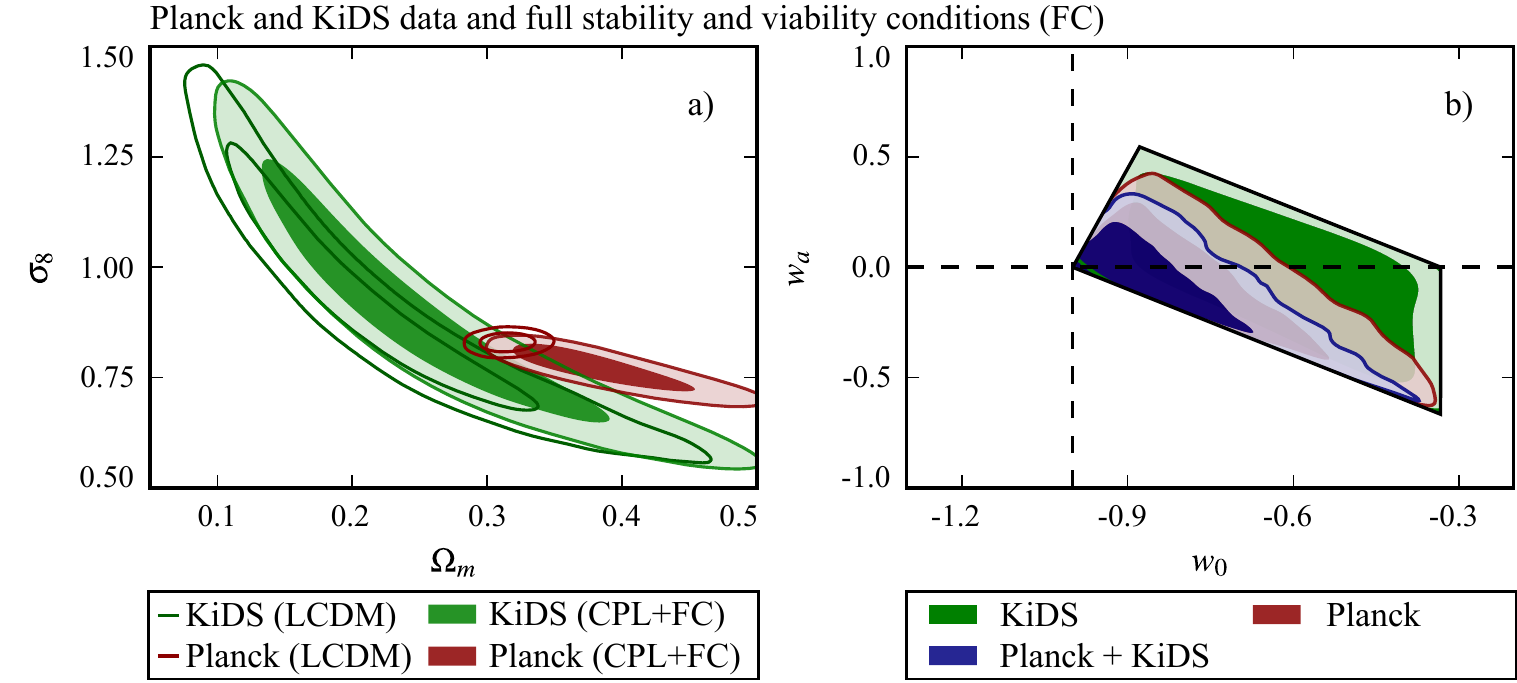}
  \caption{
	The joint marginalized posterior of $\Omega_m$ and $\sigma_8$ (panel a) and $w_0$ and $w_a$ (panel b) as obtained analyzing KiDS (red contours) and Planck (blue contours) data. Filled contours refer to constraints obtained in CPL with FC, while empty contours (left panel only) refer to $\Lambda$CDM constraints.
  The darker and lighter shades correspond respectively to the $68\%$ C.L. and the $95\%$ C.L. regions. The dashed line indicates the point corresponding to the $\Lambda$CDM model in the $w_0-w_a$ plane.  
 }
  \label{fig:fullres}
 \end{center}
\end{figure*}

The left panel of Figure~\ref{fig:fullres} shows instead the constraints achieved when FC are used in the analysis. 
As discussed above, in this case the Planck data prefer higher values of $\Omega_m$. Even though the tension with KiDS data on the $S_8$ parameter is eased with respect to the $\Lambda$CDM case ($T(S_8)=1.3$), the concordance between the two datasets is worsened ($\log{\mathcal{I}}=-0.76$).
As shown in the right panel of Figure~\ref{fig:fullres}, in this case the constraints on $w_0$ and $w_a$ are compatible with $\Lambda$CDM $w=-1$;
additionally, as can be seen in Table~\ref{tab:modsel}, the CPL model is disfavored with respect to $\Lambda$CDM when we account for FC
($\Delta DIC=4.6$ when the two datasets are combined) because the fit does not improve significantly over the $\Lambda$CDM one while the parameter space dimension grows.
This is due to the fact that the $w(z)<-1.0$ region, which is the one favoured by the data in PPF and MC analysis, is here completely 
removed by the FC, as physically viable single field quintessence models are not able to reproduce this evolution.

\begin{table}
\begin{center}
\begin{tabular}{|c|c|c|}
\hline
              &  $ T(S_8) $      & $\log{\mathcal{I}}$ \\
\hline 
$\Lambda$CDM  &  $ 2.3\sigma $          & $ -0.48 $     \\
\hline
CPL + MC  &  $ 1.0\sigma $          & $ 0.97 $     \\
\hline
CPL + FC  &  $ 1.3\sigma $          & $ -0.76 $     \\
\hline
\end{tabular}
\caption{Tension ($T$) and concordance ($\log{\mathcal{I}}$) parameters between KiDS and
Planck data in $\Lambda$CDM and in CPL with MC and with FC.}
\label{tab:tension} 
\end{center}
\end{table}

\begin{table}
\begin{center}
\begin{tabular}{|c|c|c|}
\hline
             & $\Delta \chi^2_{\rm eff}$ & $\Delta$DIC \\
\hline
\multicolumn{3}{|c|}{CPL + MC}    \\
\hline
  KiDS        & $ -0.02 $            & $  3.2 $ \\
  Planck      & $ -2.9 $            & $ -2.0 $ \\
  Planck+KiDS & $ -5.4 $            & $ -5.4 $ \\
\hline
\hline
\multicolumn{3}{|c|}{CPL + FC}    \\
\hline
  KiDS        & $ -0.3 $            & $ 0.2 $ \\
  Planck      & $  1.6 $            & $ 3.2 $ \\
  Planck+KiDS & $  1.0 $            & $ 4.6 $ \\
\hline
\end{tabular}
\caption{Model comparison through the obtained values of $\Delta \chi^2_{\rm eff}$ and $\Delta$DIC using as
a reference $\chi^2_{\rm eff}$ the one obtained in a $\Lambda$CDM analysis.}
\label{tab:modsel} 
\end{center}
\end{table}

\section{Conclusions}\label{sec:conc}
In this work we revisited the interesting possibility of easing the tension between weak lensing and 
CMB data with a dynamical dark energy, whose equation of state is described by the CPL parametrization, in light of the results of~\citep{Joudaki:2016kym}. In particular, we explored whether the model favored by Planck and KiDS data in~\citep{Joudaki:2016kym} could correspond to a theoretically viable single field quintessence, by restricting the parameter space of CPL to that corresponding to \emph{stable standard quintessence}. With the latter, we indicate DE models corresponding to one scalar field minimally coupled to gravity and without higher order derivative operators.  This theoretical assumption leads to restrictions on the
allowed  parameter space of the CPL parametrization, as described in Section~\ref{sec:stab}. 
In order to study the impact of these conditions on the observational constraints,
we performed an analysis analogous to the one done in~\citep{Joudaki:2016kym}, using
the theoretical conditions of stability and viability built-in  in \texttt{EFTCAMB}.

We considered two types of conditions: Mathematical (MC) and the 
Physical Stability Conditions (PC). The former are rather generic and consider only
the numerical stability of the model, without any theoretical consideration. 
The latter are more restrictive, ruling out all models
with ghost or gradient instabilities on the basis of theoretical considerations~\citep{Carroll:2003PhRvD, Creminelli:2009JCAP}. 
One convenient way to compare these two classes of conditions is to look at the corresponding parameter space in the $(w_0, w_a)$-plane.
While MC allow for the phantom divide crossing, 
the full set, FC, strictly forbids the region corresponding to $w<-1$, since for smooth single field quintessence  ghost instabilities would develop. These instabilities could be avoided with the inclusion of higher order derivative operators or additional degrees of freedom in the dark sector, but in both cases we would move away from single field, relatively smooth quintessence~\citep{Creminelli:2009JCAP,2005PhRvD..71d7301H}.

After performing a fit to KiDS and Planck data by means of \texttt{EFTCosmoMC}, we find that in the MC case, even though the allowed parameter space is shrunk with respect to 
the PPF case, the expansion history found in~\citep{Joudaki:2016kym} and able to ease the $S_8$
tension is still allowed by the implemented conditions. Therefore,
CPL under MC conditions in our implementation yields results analogous to the general PPF case. In particular we find $\Delta DIC=-5.4$, showing a moderate preference of the data for the model. 
Interestingly, the result changes when we apply the FC: in this case, the parameter space of CPL is significantly reduced and
both KiDS and Planck contours move towards  higher values of $\Omega_m$, as shown in Fig.~\ref{fig:allstabres}, 
with the net effect of decreasing the tension parameter to $T(S_8) = 1.3 \sigma$.
However, the value of $\Delta DIC=4.6$ (for Planck + KiDS) shows that the model is 
disfavored with respect to $\Lambda$CDM. 

In conclusion, as expected, the CPL expansion history favoured by Planck+KiDS data per~\citep{Joudaki:2016kym}, cannot correspond to stable single field quintessence, and this is shown very clearly by the contours in Fig.~\ref{fig:fullres}, where we plot the outcome of our analysis including stability conditions. The best fit model of~\citep{Joudaki:2016kym} shall rather  be interpreted in terms of multiple fields  scenarios, with the DE d.o.f. remaining relatively smooth with respect to dark matter.
The results we obtained highlight how parametrized approaches to $\Lambda$CDM extensions, although extremely useful to agnostically explore departures
from the standard cosmological model, need to be complemented with theoretically priors if one wants to connect the results to viable models of DE.
Our method allows us to quantify the tension between data sets and the preference of models while restricting to specific classes of models in a statistically meaningful way. In this case, we find that,  if one indeed restricts to standard quintessence and invokes conditions of theoretical stability, $\Lambda$CDM remains a better fit to the data. We leave for future work the exploration of the tension between Planck and KiDS data sets in more general models of dark energy, e.g. stable and viable Horndeski models.

\section*{Acknowledgments}
We thank Noemi Frusciante, Wayne Hu, Massimo Viola for useful discussions and comments, and Shahab Joudaki for clarifications about the implementation of KiDS dataset. SP and AS acknowledge support from The Netherlands Organization for Scientific Research (NWO/OCW), and from the D-ITP consortium, a program of the Netherlands Organisation for Scientific Research (NWO) that is funded by the Dutch Ministry of Education, Culture and Science (OCW). MM acknowledge support by the Foundation for Fundamental Research on Matter (FOM) and the Netherlands Organization for Scientific Research / Ministry of Science and Education (NWO/OCW).
MR is supported by U.S. Dept. of Energy contract DE-FG02-13ER41958.

\bibliographystyle{aipnum4-1}
\bibliography{bibliography}

\end{document}